\documentclass[10pt, twocolumn]{article}
\usepackage{amsmath,amssymb,amsthm}
\usepackage{bbm}
\usepackage{psfrag}
\usepackage{graphicx}
\usepackage[bookmarks=false]{hyperref}
\usepackage{authblk}
\usepackage{color}

\title{\LARGE \bf
Green Base Station Placement for Microwave Backhaul Links
}

\author[1]{Alonso Silva\thanks{Email: \href{mailto:alonso.silva@nokia-bell-labs.com}{alonso.silva@nokia-bell-labs.com}
To whom correspondence should be addressed.}}
\author[2]{Antonia Maria Masucci\thanks{Email: \href{mailto:antoniamaria.masucci@orange.com}{antoniamaria.masucci@orange.com}}}
\affil[1]{Nokia Bell Labs\\ Centre de Villarceaux\\ Route de Villejust\\ 91620 Nozay\\ France}
\affil[2]{Orange Labs\\ 44 Avenue de la R\'epublique, 92320 Ch\^atillon, France}
\date{}

\begin{document}

\maketitle
\thispagestyle{empty}
\pagestyle{empty}

\begin{abstract}
Wireless mobile backhaul networks have been proposed as a substitute in cases in which wired alternatives are not available due to economical or geographical reasons.  In this work, we study the location problem of base stations in a given region where mobile terminals are distributed according to a certain probability density function and the base stations communicate through microwave backhaul links.  Using results of optimal transport theory, we provide the optimal asymptotic distribution of base stations in the considered setting by minimizing the total power over the whole network.
\end{abstract}

\section{Introduction}

There are several scenarios in which wired alternatives are not the best solution to satisfy the traffic demand of users
due to economical or geographical reasons.
Wireless mobile backhaul networks have been proposed as a solution to these types of situations.
Since these networks do not require costly cable constructions,
they reduce total investment costs.
However, achieving high speed and long range in wireless backhaul networks remains
a significant technical challenge.

In this work, we consider a clique network of base stations and
we assume that data is transmitted independently
in different radio frequency channels.
We use optimal transport theory, also known as theory of mass transportation,
to determine in this simplified scenario
the optimal asymptotic placement of base stations
communicating through backhaul links.

Optimal transport theory has its origins
in planning problems, where a central planner
needs to find a transport plan between
two non-negative probability measures
which minimizes the average transport cost.
Resource allocation problems and/or assignment problems coming from engineering or economics
are common applications of this theory.

In the present work,
we study the problem of minimizing the total power used by
the network to achieve a certain throughput
and we use recent results of optimal transport theory to find
the optimal asymptotic base stations locations.

\subsection{Related Works}

Location games have been introduced by Hotelling~\cite{hotelling},
who modeled the spatial competition along a street
between two firms for persuading the largest number
of customers which are uniformly distributed.
Problems similar to location games,
as for example the maximum capture problem,
have been analyzed by \cite{plastria,gabszewicz} and references therein.

Within the communication networks community, Altman et al.~\cite{spatial,AltmanKSS12} studied the duopoly situation in the uplink scenario of a cellular network
where users are placed on a line segment. 
Considering the particular cost structure that arises in the cellular context, 
the authors observe that complex cell shapes are obtained at equilibrium.
Silva et al.~\cite{silva2013,SilvaTAD10a,SilvaTAD10b} analyzed the problem of mobile terminals association to base stations
using optimal transport theory and considering the data traffic congestion produced by this mobile terminals to base stations association.

\subsection{Energy efficiency}

Our objective is to conceive wireless backhaul networks
able to guarantee quality of service while
minimizing the energy consumption of the system.
We follow the free space path loss model
in which the signal strength drops in 
proportion to the square of the distance
between transmitter and receiver since it
is a good approximation for outdoor scenarios.

The works on stochastic geometry are related
to our study (see e.g. the books of Baccelli and Blaszczyszyn~\cite{baccelli}, \cite{baccelli2})
but we do not consider any particular deployment distribution function such as e.g. Poisson point processes.

The remaining of this work is organized as follows.
In Section~\ref{sec:model} we provide the model formulation of the considered problem
where we redefine the probability density function of mobile terminals to incorporate
their throughput requirements and determine the power cost function of inter-cell and intra-cell
communications.
In Section~\ref{sec:results} we provide the main results of our work
by considering the free-path loss approximation and asymptotic results
from optimal transportation theory.
In Section~\ref{sec:simulations} we provide illustrative simulations for the asymptotic results
obtained in the previous section,
and in Section~\ref{sec:conclusions} we conclude our work.

\section{Model formulation}\label{sec:model}

A summary of the notation used on this work
can be found in Table~\ref{notation}.

\begin{table}
\renewcommand{\arraystretch}{1.8}
\caption{Notation}
\label{notation}
\centering
\begin{tabular}{|c|l|}
    \hline
    $N$ & Total number of mobile terminals in the network\\ \hline
    $K$ & Total number of base stations\\ \hline
    $f$ & Deployment distribution function of mobile terminals\\ \hline
    $(x_k,y_k)$ & Position of the $k$-th base station\\ \hline
    $C_i$ & Cell determined by the $i$-th base station\\ \hline
    $N_i$ & Number of mobile terminals associated to the $i$-th BS\\ \hline
    $h_i$ & Channel gain function in the $i$-th cell\\ \hline
	$h_{ij}$ & Channel gain between base station $i$ and base station $j$\\ \hline
	$m_i$ & Traffic requirement satisfied by base station $i$\\ \hline
	$m$ & Total traffic requirement satisfied by the network\\ \hline
\end{tabular}
\end{table}
We are interested on the analysis of a microwave backhaul network
deployed over a bounded region,
which we denote by~$\mathcal{D}$,
over the two-dimensional plane.
Mobile terminals are distributed
according to a given probability density function~\mbox{$f(x,y)$}.
The proportion of mobile terminals in a sub-region
\mbox{$\mathcal{A}\subseteq\mathcal{D}$} is
\begin{equation*}
\int\!\!\!\!\int_\mathcal{A} f(x,y)\,dx\,dy.
\end{equation*}

The number of mobile terminals in sub-region
\mbox{$\mathcal{A}\subseteq\mathcal{D}$}, denoted by~$N(\mathcal A)$,
can be approximated by
\begin{equation*}
N(\mathcal A)=N\cdot\left(\int\!\!\!\!\int_\mathcal{A} f(x,y)\,dx\,dy\right),
\end{equation*}
where~$N$ denotes the total number of mobile terminals in the network.

We consider $K$~base stations in the network, denoted by $\mathrm{BS}_1, \mathrm{BS}_2, \ldots, \mathrm{BS}_K$,
at positions \mbox{$(x_1,y_1), (x_2,y_2),\ldots, (x_K, y_K)$} to be determined.
Our objective is to minimize
the energy consumption in the system.

We denote by~$C_i$ the set of mobile terminals associated to base station $\mathrm{BS}_i$
and by~$N_i$ the number of mobile terminals within that cell, i.e., the cardinality of the set~$C_i$.

\subsection{Modification of the distribution function}\label{subsec:modification}

The probability density function
and the throughput requirements of mobile terminals
both depend on the location.
To simplify the problem resolution,
we consider the following modification of the probability density function to have
the location dependency in only one function.
The probability density function of mobile terminals considered in our work and denoted by $f(x,y)$ is general. 
Instead of considering a particular probability density function, 
denoted by $\tilde f(x,y)$, and an average throughput requirement,
denoted by $\tilde\theta(x,y)$,
in each location~$(x,y)$,
we consider a constant throughput~$\theta>0$ to be determined and redefine the distribution of mobile terminals~$f(x,y)$ as
follows
\begin{equation*}
f(x,y):=\frac{\tilde f(x,y)\tilde\theta(x,y)}{\theta}\quad\textrm{for all }(x,y)\in\mathcal{D}.
\end{equation*}
Since $f(x,y)$ must be a probability density function, we need to impose
\begin{equation*}
\int\!\!\!\!\int_\mathcal{D} f(x,y)\,dx\,dy=1,
\end{equation*}
or equivalently,
\begin{equation*}
\frac{1}{\theta}\int\!\!\!\!\int_\mathcal{D} \tilde f(x,y)\tilde\theta(x,y)\,dx\,dy=1.
\end{equation*}
For this equation to hold, we have to impose
\begin{equation*}
\theta=\int\!\!\!\!\int_\mathcal{D}\tilde f(x,y)\tilde\theta(x,y)\,dx\,dy.
\end{equation*}

We have that the following equation holds:
\begin{equation*}
f(x,y)\theta=\tilde f(x,y)\tilde\theta(x,y).
\end{equation*}
The previous equation simply states that, e.g.,
a mobile terminal with double demand than another mobile terminal
would be considered as two different mobile terminals
both at the same location
with the same demand as the other mobile terminal.

Since in a microwave backhaul network,
we need to consider the energy from within
base stations and from base stations to
mobile terminals, we need to consider
both the intra-cell costs and the inter-cell costs.
This is the subject of the
following two subsections.

\subsection{Intra-cell costs}

The power transmitted, denoted by $P^T$, from base station~$\mathrm{BS}_i$
to a mobile terminal located at position~$(x,y)$ is denoted by~$P^T_i(x,y)=P_i(x,y)$.
The received power, denoted by $P^R$, at the mobile terminal located at position~$(x,y)$
associated to base station~$\mathrm{BS}_i$ is given by
$P^R_i(x,y)=P_i(x,y) h_i(x,y)$,
where $h_i(x,y)$ is the channel gain between base station~$\mathrm{BS}_i$
and the mobile terminal located at position~$(x,y)$,
for every \mbox{$i\in\{1,\ldots,K\}$}.

We assume that neighboring base stations transmit their signals in
orthogonal frequency bands and that
interference between base stations that are
far from each other is negligible.
Consequently, instead of considering the $\mathrm{SINR}$ (Signal to Interference plus Noise Ratio),
we consider as performance measure the~$\mathrm{SNR}$ (Signal to Noise Ratio).

The~$\mathrm{SNR}$ received at mobile terminals at position~$(x,y)$ in cell~$C_i$ is given by
\begin{equation*}
\mathrm{SNR}_i(x,y)=\frac{P_i(x,y) h_i(x,y)}{\sigma^2},
\end{equation*}
where~$\sigma^2$ is the expected noise power.
We assume that the associated instantaneous mobile throughput 
is given by the following expression, which is based
on Shannon's capacity theorem:
\begin{equation*}
\theta_i(x,y)=\log(1+\mathrm{SNR}_i(x,y)).
\end{equation*}

The throughput requirement translates into
\begin{equation*}
\theta_i(x,y)\ge\theta.
\end{equation*}
Thanks to our previous development,
we can consider a constant throughput requirement
through the modification of the probability density function.

Therefore, the throughput requirement becomes
\begin{equation*}
\log\left(1+\frac{P_i(x,y) h_i(x,y)}{\sigma^2}\right)=\theta,
\end{equation*}
or equivalently,
\begin{equation}\label{eq:uganda1}
P_i(x,y)=\frac{\sigma^2}{h_i(x,y)} (2^\theta-1).
\end{equation}

Therefore the intra-cell power required by base station $i$ is given by
\begin{equation}\label{eq:pascua}
P_i^{\rm intra}=\int\!\!\!\!\int_{C_i}P_i(x,y)f(x,y)\,dx\,dy.
\end{equation}

The previous equation provide us
an energy cost function for the intra-cell requirements
of the network.
In the following subsection,
our analysis will be focused
on the inter-cell requirements.

\subsection{Inter-cell costs}\label{sec:routing}

In order to take into account the routing cost,
we consider the power transmitted $P^T$ from base station~$\mathrm{BS}_i$
to base station~$\mathrm{BS}_j$ denoted by
$P^T_{ij}=P_{ij}$.
The received power $P^R$ at the receiving base station~$\mathrm{BS}_j$
from the transmitting base station~$\mathrm{BS}_i$ is given by
$P^R_{ij}=P_{ij}h_{ij}$,
where $h_{ij}$ is the channel gain between base station~$\mathrm{BS}_i$
and base station~$\mathrm{BS}_j$.
The~$\mathrm{SNR}$ received at the receiving base station~$\mathrm{BS}_j$
from the transmitting base station~$\mathrm{BS}_i$ is given by
\begin{equation*}
\mathrm{SNR}_{ij}=\frac{P_{ij} h_{ij}}{\sigma^2},
\end{equation*}
where~$\sigma^2$ is the expected noise power.
We assume that the associated instantaneous base station throughput
at the receiving base station~$\mathrm{BS}_j$ from the transmitting base station~$\mathrm{BS}_i$ 
is given by the following expression, which is based
on Shannon's capacity theorem:
\begin{equation*}
\theta_{ij}=\log(1+\mathrm{SNR}_{ij}).
\end{equation*}

Let us define by $m_i$ the traffic requirement concentrated at base station $\mathrm{BS}_i$, i.e.
\begin{equation*}
m_i=\theta\int\!\!\!\!\int_{\mathcal{C}_i} f(x,y)\,dxdy.
\end{equation*}

We assume that the traffic requirement $m_i$ concentrated at base station $\mathrm{BS}_i$
is sent at the other base stations proportionally to the traffic requirement at the other base stations.
Therefore, the traffic
between the receiving base station~$\mathrm{BS}_j$ and the transmitting base station~$\mathrm{BS}_i$
is given by $m_i(m_j/m)$.

We make the simplifying assumption \mbox{$\log(1+\mathrm{SNR}_{ij})\approx\mathrm{SNR}_{ij}$}.
Then the throughput requirement translates into
\begin{equation*}
\frac{P_{ij} h_{ij}}{\sigma^2}=m_i\frac{m_j}{m},
\end{equation*}
or equivalently
\begin{equation}\label{eq:timbuktu1}
P_{ij}=\frac{\sigma^2}{h_{ij}}\frac{m_i m_j}{m}.
\end{equation}

The power cost to transmit the traffic $m_i$ is thus given by
\begin{equation}\label{eq:peru}
\sum_{j=1}^K \frac{\sigma^2}{h_{ij}} \frac{m_i m_j}{m},
\end{equation}
where
\begin{align*}
m&=\sum_{j=1}^K m_j\nonumber\\
&=\theta\sum_{j=1}^K\int\!\!\!\!\int_{\mathcal{C}_j} f(x,y)\,dxdy\nonumber\\
&=\theta\int\!\!\!\!\int_\mathcal{D} f(x,y)\,dxdy.
\end{align*}

Similar to eq.~\eqref{eq:pascua} of the previous subsection, equation \eqref{eq:peru} provides us
a power cost function for the inter-cell requirements
of the network. In the next section,
we consider both inter-cell and intra-cell cost functions
to determine the total power cost and
obtain the asymptotic location of base stations to minimize this total power cost.

\section{Results}\label{sec:results}

From the previous section,
the total power of the network is equal to 
the sum of intra-cell power (the sum of the power used within each cell in the network)
and the inter-cell power (the sum of the power used over the pairs of communicating base stations in the network),
i.e.
\begin{equation*}
P_{\rm total}=\sum_{i=1}^K P_i^{\rm intra}+\sum_{i=1}^K\sum_{\substack{j=1 \\j\ne i}}^K P_{ij}^{\rm inter},
\end{equation*}
where
\begin{equation*}
P_i^{\rm intra}=\int\!\!\!\!\int_{C_i}P_i(x,y)f(x,y)\,dx\,dy,
\end{equation*}
is the intra-cell power consumption in cell $C_i$
and from eq.~\eqref{eq:uganda1}, we obtain
\begin{equation*}
P_i^{\rm intra}=\int\!\!\!\!\int_{C_i} \frac{\sigma^2}{h_i(x,y)} (2^\theta-1)f(x,y)\,dx\,dy,
\end{equation*}
and from eq.~\eqref{eq:timbuktu1} the inter-cell power consumption is
\begin{equation*}
P_{ij}^{\rm inter}=\frac{\sigma^2}{h_{ij}}\frac{m_i m_j}{m}.
\end{equation*}

In the following subsection,
thanks to the free-space path loss approximation,
we are able to find an expression for the channel gain
and the total power cost.

\subsection{Free-space path loss approximation}

Let $d_i(x,y)$ denote the Euclidean distance
between mobile terminal at position~$(x,y)$ and base station~$\mathrm{BS}_i$
located at~$(x_i,y_i)$, i.e.
\begin{equation*}
d_i(x,y)=\sqrt{(x_i-x)^2+(y_i-y)^2}.
\end{equation*}
Similarly, let $d_{ij}$ denote the Euclidean distance
between base station~$\mathrm{BS}_i$ located at~$(x_i,y_i)$
and base station~$\mathrm{BS}_j$ located at~$(x_j,y_j)$, i.e.
\begin{equation*}
d_{ij}=\sqrt{(x_i-x_j)^2+(y_i-y_j)^2}.
\end{equation*}

The free-space path loss approximation gives us that
the channel gain between base station~$\mathrm{BS}_i$
and the mobile terminal located at position~$(x,y)$
is given by
\begin{equation*}
h_i(x,y)=d_i(x,y)^{-2},
\end{equation*}
and analogously the free-space path loss approximation give us
that the channel gain between base station~$\mathrm{BS}_i$ and base station~$\mathrm{BS}_j$
is given by
\begin{equation*}
h_{ij}=d_{ij}^{-2}.
\end{equation*}

The total power cost is therefore given by
\begin{equation}\label{eq:doumbodo1}
\sum_{i=1}^K\int\!\!\!\!\int_{C_i} (2^\theta-1)\sigma^2 d_i(x,y)^2 f(x,y)\,dx\,dy
+\frac{\sigma^2}m\sum_{i=1}^K\sum_{\substack{j=1\\j\ne i}}^K m_im_jd_{ij}^2.
\end{equation}

When the number of base stations $K$ is very large,
in our setting tends to infinity,
instead of looking at the locations of base stations $(x_i,y_i)$,
we will look into the limit density $\nu$ of the locations $(x_i,y_i)$.
To do that, we identify each set of $K$ points with the measure
\begin{equation*}
\nu_K=\frac1K\sum_{i=1}^K \delta_{(x_i,y_i)},
\end{equation*}
where $\delta_{(x_i,y_i)}$ is the delta Dirac function at location~$(x_i,y_i)$.

The asymptotic analysis of these functions has been performed (see e.g.~\cite{Bouchitte,buttazzo2005})
within the context of optimal transport theory
with the extensive use of $\Gamma$-convergence.

The inter-cell power cost in terms of the measure $\nu_K$ is given by
\begin{equation*}
\frac{\sigma^2}m\int_\mathcal{D}\int_\mathcal{D}\lVert (x_i,y_i)-(x_j,y_j)\rVert^2\,d\nu_K\,d\nu_K,
\end{equation*}
which is equivalent to
\begin{equation*}
\frac{\sigma^2}m\int_{\mathcal{D}\times\mathcal{D}}\lVert (x_i,y_i)-(x_j,y_j)\rVert^2\,d(\nu_K\otimes\nu_K).
\end{equation*}
We notice that the discrete sum of eq.~\eqref{eq:doumbodo1} becomes
an integral by considering the limit of measures $\{\nu_K\}_{K\in\mathbb{N}}$.

The total cost taking into account both cost functions gives the problem
\begin{align*}
&\operatorname*{Min}\sum_{i=1}^K\int\!\!\!\!\int_{C_i}(2^\theta-1)\sigma^2\lVert (x,y)-(x_i,y_i)\rVert^2 f(x,y)\,dxdy\nonumber\\
&+\frac{\sigma^2}m\int_{\mathcal{D}\times\mathcal{D}}\lVert (x_i,y_i)-(x_j,y_j)\rVert^2\,d(\nu_K\otimes\nu_K).
\end{align*}

We denote the function
\begin{equation*}
V(x)=\lvert x\rvert^2.
\end{equation*}

The necessary conditions of optimality (see~\cite{buttazzo2013}) are 
\begin{equation}\label{eq:toty}
(2^\theta-1)\sigma^2\phi+\frac{2\sigma^2}{m}V*\nu=c\quad\nu-\textrm{a.e.}
\end{equation}
where $\phi$ is the Kantorovich potential for the transport from~$f$
to $\nu$ and $c$ is the Lagrange multiplier of the mass constraint on $\nu$.

A connection between the Kantorovich potential $\phi$ and the transport map $T$
from $f$ to $\nu$ is given by the Monge-Amp\`ere equation
\begin{equation*}
f=\nu(T)\,\mathrm{det}(\nabla T).
\end{equation*}

From equation~\eqref{eq:toty}, we obtain
\begin{equation*}
(2^\theta-1)\sigma^2\nabla\phi+\frac{2\sigma^2}{m}\nabla V*\nu=0.
\end{equation*}

Since
\begin{equation*}
T(x)=x-\nabla\phi(x).
\end{equation*}
Therefore we have the system
\begin{equation}\label{eq:philippe}
\left\{
\begin{array}{l}
(2^\theta-1)\sigma^2(x-T(x))+\frac{2\sigma^2}{m}\nabla V*\nu=0\\
f=\nu(T)\mathrm{det}(\nabla T).
\end{array}
\right.
\end{equation}

We can proceed by an iterative scheme,
fixing an initial $\nu_0$ and obtaining $T_0$
from the first equation of the system~\eqref{eq:philippe}
and obtaining $\nu_1$ from the second equation
and proceed iterating the scheme above.

The previous system of equations allows us
to find the optimal asymptotic base stations placement~$\nu$
as a function of the distribution of mobile terminals~$f$
when the solution exists.

\section{Simulations}\label{sec:simulations}

We follow the development done in~\cite{buttazzo2013}.
We simulate the example for the one-dimensional case.
If we suppose that the barycenter of $\nu$ is in the origin,
we obtain:
\begin{equation*}
V*\nu=mx^2+\int y^2\,d\nu(y),
\end{equation*}
so that
\begin{equation*}
A\phi'(x)+4Bx=0,
\end{equation*}
which gives
\begin{equation*}
\phi'(x)=-\frac{4}{(2^\theta-1)} x
\quad\textrm{and}\quad
T(x)=\left(1+\frac{4}{(2^\theta-1)}\right).
\end{equation*}
Putting previous expressions in the one-dimensional Monge-Amp\`ere
equation and indicating by $v$ the density of $\nu$, we obtain
\begin{equation*}
f(x)=\nu\left(\left(1+\frac{4}{(2^\theta-1)}\right)x\right)\left(1+\frac{4}{(2^\theta-1)}\right),
\end{equation*}
and changing variables
\begin{equation}\label{eq:uganda}
v(y)=\frac{1}{1+\frac{4}{(2^\theta-1)}}f\left(\frac{y}{1+\frac{4}{(2^\theta-1)}}\right).
\end{equation}

\begin{figure}[h!]
  \centering
    \includegraphics[width=0.5\textwidth]{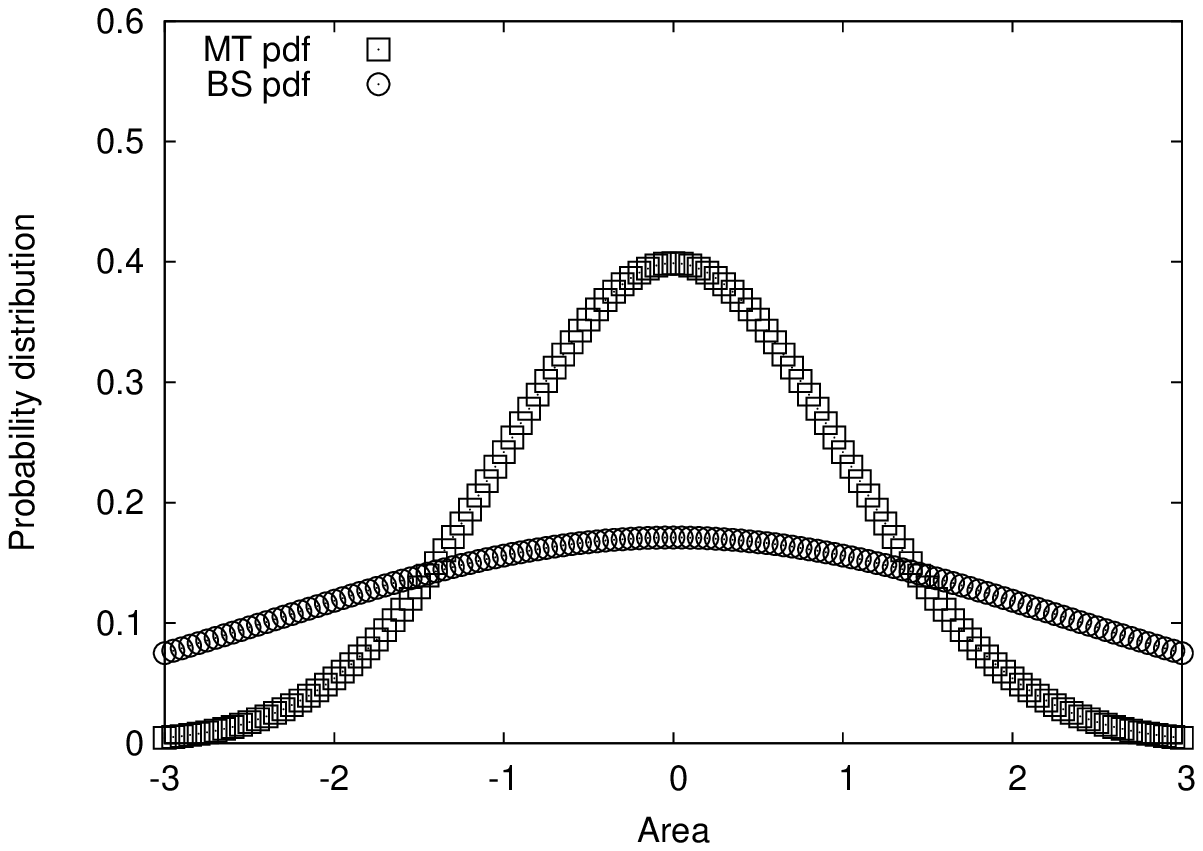}
  \caption{Optimal probability distribution function of the base stations given the probability density function
of the mobile terminals.}
\label{fig:pdfbs2}
\end{figure}

\begin{figure}[h!]
  \centering
    \includegraphics[width=0.5\textwidth]{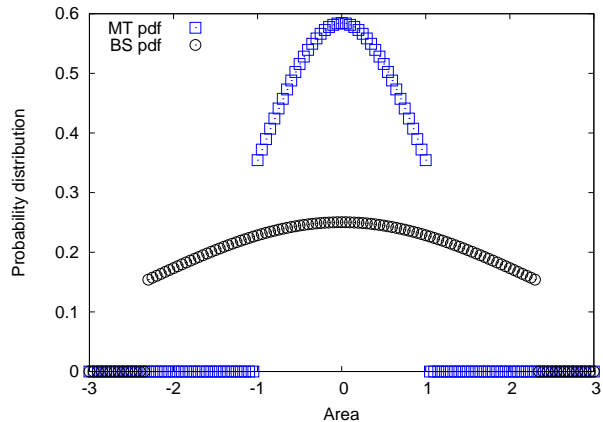}
  \caption{Optimal probability distribution function of the base stations given the probability density function
of the mobile terminals.}
\label{fig:pdfbs}
\end{figure}

We consider that mobile terminals are distributed over the line
as a normal distribution function with zero mean and
standard deviation equal to one, i.e. $\mathcal{N}(0,1)$.
We consider that the throughput requirement $\theta$ is constant and equal to $24$~\textrm{Kbps}.
We notice that as explained in subsection~\ref{subsec:modification}
we could have considered a non-constant throughput requirement and
redefine the mobile terminal probability density function for the throughput requirement to be constant.
From eq.~\eqref{eq:uganda}, we can compute the optimal base station distribution given by Fig.~\ref{fig:pdfbs2}.
We notice that the optimal base station distribution corresponds to a smoother normal probability distribution.

Motivated by the previous simulation, we consider a second scenario where mobile terminals are distributed
as a truncated normal distribution function between~$[-1,1]$.
From eq.~\eqref{eq:uganda}, we can compute the optimal base station distribution given by Fig.~\ref{fig:pdfbs}.
From Fig.~\ref{fig:pdfbs}, we notice 
that surprisingly the optimal base stations distribution has a support that does not coincide with the
mobile terminals distribution.
If we only consider the intra-cell cost we would have obtained the exact same probability distribution
of mobile terminals (it is easy to see since in that case the cost would have been zero).
We have thus verified that the optimal base station
probability density function would have been modified
by the intra-cell cost as given by Fig.~\ref{fig:pdfbs}.
Similar to the first scenario, the optimal base station distribution corresponds
to a smoother probability distribution.

\section{Conclusions}\label{sec:conclusions}

In this work, we investigated the asymptotic optimal placement of
base stations with microwave backhaul links. We considered
the problem of minimizing the total power of the network
while maintaining a required throughput. Using optimal transport theory,
we provided the optimal asymptotic base station placement.
Moreover, the case where routing cost is taken into account is also analyzed.

\section*{Acknowledgment}
The work of A.~Silva has been partially carried out at LINCS (\url{http://www.lincs.fr}).
\bibliographystyle{hieeetr}
\bibliography{mybibfile}
\end{document}